\documentstyle[aps,preprint]{revtex}

\def\be{\begin{eqnarray}}
\def\ee{\end{eqnarray}}

\def\pragueF{Presented at the {\bf Fifth Central-European Workshop on Quantum
Optics}, Prague, Czech Republic, April 25 - 28, 1997}

\begin{document}

\draft

\title{REALIZATION OF NONLINEAR OSCILLATORS \\ WITH A TRAPPED ION
\footnote{\pragueF}}

\author{G. Drobn\'y$^{(a)}$ and B. Hladk\'y$^{(b)}$}

\address{
$^{(a)}$ Institute of Physics, Slovak Academy of Sciences, \\
           D\'ubravsk\'a cesta 9, 842 28 Bratislava, Slovakia;
           e-mail: drobny@savba.sk \\
$^{(b)}$Department of Optics, Comenius University,  Mlynsk\'a dolina,
           \\ 84215 Bratislava, Slovakia; e-mail: hladky@fmph.uniba.sk}

\date{April 25, 1997}

\maketitle

\abstract{
We consider a trapped ion with a quantized center--of--mass
motion in 2D trap potential.
With external laser fields the effective (non)linear coupling
of two orthogonal vibrational modes can be established via
stimulated Raman transition.
Nonclassical vibrational states such as squeezed states or
two--mode entangled states (Schr\"odinger cat--like states)
can be generated.
When the vibrational modes are entangled with internal energy levels
the Greenberger--Horne--Zeilinger (GHZ) states can be prepared.
}

\section{Motivation}

The Jaynes--Cummings model (JCM) \cite{Jaynes 1963} which
describes an interaction of one two--level atom
with a quantized cavity field belongs to the fundamentals
of quantum optics.
The successful experimental realization of the JCM is associated
with Rydberg atoms in a high-Q microwave cavity.
The quantum effects such as the collapse--revival behavior and
preparation of Schr\"odinger cat--like states
have been demonstrated \cite{Brune 1996}.

Recent experimental developments in laser cooling and ion trapping
\cite{Diedrich 1989}
have enabled to realize a formal analogue of the JCM
in which the cavity field mode is replaced by a quantized vibrational
mode of the center--of--mass motion of a trapped ion
\cite{Blockley 1992,Monroe 1996,Vogel 1995}.
There are two virtues in experiments with trapped ions.
Firstly, dissipative effects which are inevitable from cavity
damping in the microwave or optical domain can be significantly
suppressed for the ion motion. Secondly, the motion of the trapped ion
can be well controlled by proper sequences of laser pulses
tuned either to the atomic electronic transition or to resolved
vibrational sidebands. These aspects make trapped ions
candidates for quantum computing \cite{Cirac 1995}.
It is worth noticing that
the most prominent nonclassical vibrational states
of a trapped ion (coherent, squeezed, Fock and
Schr\"odinger cat--like states)
have been successfully prepared in laboratory
\cite{Monroe 1996}.

The formal analogy of the vibrational mode
of a trapped ion with the cavity field mode can be extended
to multi--mode systems. In particular, schemes for preparing
correlated two--mode Schr\"odinger cat states, the Bell  and
SU(2) states of vibrational motion in a two--dimensional trap
have been proposed \cite{Gou 1996}.

In this paper we consider a trapped ion with a quantized center--of--mass
motion in 2D trap potential.
Irradiating the ion with two external laser fields,
which are tuned to well--resolved vibrational sidebands
to stimulate Raman transition between internal energy levels,
the effective linear or nonlinear coupling of two orthogonal vibrational
modes can be established.
It is shown how to realize with a trapped ion also analogues
of other fundamental elements of quantum optics -
nonlinear optical processes (multiwave mixings)
such as the degenerate two--photon down conversion.


\section{The models}

Consider a quantized center--of--mass motion of an ion
which is confined in a 2D harmonic potential
characterized by the trap frequencies $\nu_x$ and $\nu_y$
in two orthogonal directions $x$ and $y$.
The ion is irradiated by two external laser fields
with frequencies $\omega_x$, $\omega_y$ along the $x$ and $y$ axes.
The laser fields stimulate transitions between three internal energy
levels $a, b, c$ in $\Lambda$ configuration (with the upper $c$ level).
The interaction Hamiltonian for the system under consideration can be
written in the form:
\be
\hat{H}_{int} &=&
\frac{1}{2}\hbar\Omega_{x} \left[
\mbox{e}^{-i\omega_{x}t} \hat{D}_{x}(i\epsilon_x) |c\rangle \langle a| +
\mbox{e}^{i\omega_{x}t} \hat{D}_{x}(-i\epsilon_x) |a\rangle \langle c|
\right] \nonumber \\
&+&\frac{1}{2}\hbar\Omega_{y} \left[
\mbox{e}^{-i\omega_{y}t} \hat{D}_{y}(i\epsilon_y) |c\rangle \langle b| +
\mbox{e}^{i\omega_{y}t} \hat{D}_{y}(-i\epsilon_y) |b\rangle \langle c|
\right]
.\label{ha1} \ee
The absorption (emission) of energy from (to)
external laser field in $q$ direction ($q=x,y$)
is accompanied by momentum exchange between the field and the ion
which is described in (\ref{ha1}) by the displacement operator
$\hat{D}_{q}(i\epsilon_q)=
\exp[i\epsilon_q(\hat{a}^{\dagger}_{q}+\hat{a}_{q})]$;
$\hat{a}^{\dagger}_{q}$, $\hat{a}_{q}$ are the creation
and annihilation operators of vibrational quanta in a given direction.
The parameter $\epsilon_q$ is defined as
$\epsilon_q^2=E_q^{(r)}/(\hbar\nu_q)$
where $E_q^{(r)}$ is the classical recoil energy of the ion;
$\Omega_q$ is a Rabi frequency of
the laser--driven transition and
$|i\rangle \langle j|$ is an atomic transition
operator. 

In the Lamb-Dicke limit $\epsilon_x \approx \epsilon_y \ll 1$
only resonant processes can be taken into account.
The upper $c$ level can be adiabatically eliminated
under resonance conditions on the frequencies of the laser fields:
\be
E_{a}/\hbar+\omega_{x}+ m \nu_{x}=
E_{b}/\hbar+\omega_{y}+ n \nu_{y}
\label{re1}\ee
(i.e., $m$ and $n$ trap quanta are involved in the stimulated
Raman transition between internal energy levels $a$ and $b$)
and off--resonance conditions for the transitions from levels
$a$ and $b$ to the upper level $c$ 
(with detuning $\Delta\gg m\nu_x, n\nu_y$):
\be
E_{c}-E_{a}=\hbar\omega_{x}+m \hbar\nu_{x}+\hbar\Delta, \qquad
E_{c}-E_{b}=\hbar\omega_{y}+n \hbar\nu_{y}+\hbar\Delta
.\ee

The effective interaction Hamiltonian in the rotating--wave-approximation
and the interaction picture reads:
\be
\hat{H}_{eff}=
  &-&\left[ \frac{\hbar(-1)^n (i\epsilon)^{m+n} \Omega_{x}\Omega_{y}}
            {4m!n!\Delta}
        \hat{a}_{x}^{m} \hat{a}_{y}^{\dagger n} |b\rangle \langle a|
        + h.c. \right]  \nonumber \\
  &-&\left[\frac{\hbar \epsilon^{2m} \Omega_{x}^2}{4m!^2\Delta}
           \hat{a}_{x}^{m} \hat{a}_{x}^{\dagger m} |a\rangle \langle a| +
           \frac{\hbar \epsilon^{2n} \Omega_{y}^2}{4n!^2\Delta}
           \hat{a}_{y}^{n} \hat{a}_{y}^{\dagger n} |b\rangle \langle b|
    \right]
\label{ha2}
.\ee
The terms in the second line represent generalized 
Stark shifts of the levels $a$ and $b$.

For example, for $m=n=1$ one gets $\hat{H}_{eff}=\hbar \lambda
(\hat{a}_x\hat{a}_y^\dagger |b\rangle \langle a| +
\hat{a}_x^\dagger \hat{a}_y |a\rangle \langle b|)$.
This exactly solvable interaction Hamiltonian is known
in the cavity QED within a context of two-photon transitions of
a two-level atom interacting with a bichromatic field in a cavity
as well for an ion in 1D trap with a resonator \cite{Gerry 1990}.
The mutual entanglement (correlation) of two vibrational
modes with ionic internal degrees of freedom is
established. When the vibrational modes $x,y$ are initially prepared in
coherent states $|\beta\rangle_x, |\gamma\rangle_y$,
the dynamics of the vibrational modes $x,y$ in phase space
is characterized by revival times
$t_R^{(x)}\approx \frac{2\pi\beta}{\lambda \gamma}$,
$t_R^{(y)}\approx \frac{2\pi\gamma}{\lambda \beta}$
(i.e., the initial quasidistributions are partially restored).
If $\beta\approx\gamma$ the revival times are independent
of the coherent amplitudes and a collapse--revival structure
of the atomic inversion can be observed.
Similarly to the JCM, the quasidistributions bifurcate
into two components.
Nevertheless, the vibrational modes are entangled
with the internal energy levels. To disentangle them,
conditional measurements on the internal energy levels
can be performed \cite{Vogel 1993}.
For example, the internal state of the ion can be determined
by driving transition from the level
$b$ to an auxiliary level $d$ and observing the fluorescence signal.
No signal (no interaction with probing field) means
the undisturbed ion in the level $a$.
One can thus prepare {\it pure} two--mode correlated states
of the vibrational system.
At half of the revival time $t_R^{(x)}\approx t_R^{(y)}$
one finds a structure close to a pure two--mode Schr\"odinger
cat--like state which consists of four components. Each particular
vibrational mode is in a two--component mixture (the components are
mutually rotated by $\pi$ in the phase space).

It is worth noticing that GHZ states of importance in tests of quantum
mechanics can be generated in a rather simple way
when the vibrational modes are still entangled with internal energy levels.
Preparing the initial state with one trap quantum in $x$ mode, i.e.,
$|\psi(0)\rangle=|1\rangle_x |0\rangle_y |a\rangle$,
one gets at quarter of the Rabi cycle $\lambda t=\pi/4$ the GHZ state
$|\psi\rangle=\frac{1}{\sqrt{2}}(|1\rangle_x |0\rangle_y |a\rangle-
i|0\rangle_x |1\rangle_y |b\rangle)$.
If instead of (\ref{re1}) the resonance condition
$E_{a}/\hbar+\omega_{x} + \nu_{x}= E_{b}/\hbar+\omega_{y} - \nu_{y}$
is assumed, one can start with vibrational modes in the
vacuum and the ion in the level $b$ to produce a GHZ state.

In the degenerate case, when the level $b$ is identical with the level $a$
all the atomic operators can be eliminated (the $c$ level is far of resonance)
and the effective interaction Hamiltonian takes the form
\be
\hat{H}_{eff}^{(xy)}=
  &-&\left[ \frac{\hbar(-1)^n (i\epsilon)^{m+n} \Omega_{x}\Omega_{y}}
                      {4m!n!\Delta}
            \hat{a}_{x}^{m} \hat{a}_{y}^{\dagger n}
    + h.c. \right]  \nonumber \\
  &-&\left[ \frac{\hbar \epsilon^{2m} \Omega_{x}^2}{4m!^2\Delta}
          \hat{a}_{x}^{m} \hat{a}_{x}^{\dagger m} +
          \frac{\hbar \epsilon^{2n} \Omega_{y}^2}{4n!^2\Delta}
          \hat{a}_{y}^{n} \hat{a}_{y}^{\dagger n}
    \right]
\label{ha3}
.\ee
For $m=n=1$ the linear coupler is obtained.
Nontrivial case - the nonlinear coupling of the bosonic modes -
is realized when one of the lasers is tuned to the second
vibrational sideband. For example $m=1$, $n=2$ leads
to two--phonon down conversion governed by
$\hat{H}_{eff}^{(xy)}=\hbar\lambda_2
(\hat{a}_{x} \hat{a}_{y}^{\dagger 2} +
 \hat{a}_{x}^{\dagger} \hat{a}_{y}^2)$.
One can thus simulate
the degenerate three--wave mixing \cite{Perina 1991}
even on the long time--scale.
It is well--known that starting with coherently
excited $x$ and empty $y$ mode the squeezed vacuum in the latter
is produced. Besides of that at the time of the maximal depletion
of the $x$ mode the oscillations in photon number distributions
of both modes can be observed. In the $y$ mode a two--component mixture
is established with a remarkable interference pattern between
the components (peaks) of the Wigner function \cite{Drobny 1997}.

For experimentalists also the case $m=1$, $n=3$ can be of interest.
In the three--phonon down conversion a three--fold structure
is formed in the phase space of the initially empty $y$ mode.
In its Wigner function strong interference structures between
three ``arms'' (rotated by $2\pi/3$ in phase space) 
are rapidly developed. The initially coherent
$x$ mode undergoes radial splitting into two components.
The resulting {\it mixture} exhibits the features typical for
{\it pure} Schr\"odinger cat states
(significant interference pattern in the Wigner function
and oscillations of photon number distribution) \cite{Drobny 1997}.

\section{Conclusion}

To conclude, the formal analogy of the vibrational mode
of a trapped ion with the cavity field mode in the JCM
leads to the possibility of realizing some cavity QED ideas
without using an optical cavity.
We showed that the trapped ion can serve in the Lamb--Dicke limit
as a simulator of nonlinear optical processes, namely multiwave mixings,
without crucial limitations on the interaction length.
The damping of bosonic vibrational modes can be significantly suppressed.
The dominant decoherence effect is the spontaneous emission of
the ion which is not taken into account here.  

\section*{Acknowledgments}
We thank V. Bu\v{z}ek for useful suggestions and comments.
This work was supported by the grant agency VEGA
of the Slovak Academy of Sciences under the project
2/1152/96.


\end{document}